\documentclass[prl,twocolumn,showpacs,floatfix,longbibliography]{revtex4-2}
\usepackage{mathrsfs,braket}
\usepackage{amssymb, amsbsy, amsmath, latexsym, dsfont, array, layout, graphicx, mathrsfs, color, ulem, bm}
\usepackage{amsthm}
\usepackage[colorlinks=true,linkcolor=blue,anchorcolor=red,citecolor=blue, urlcolor=blue]{hyperref}

\begin{document}
\title{Universal Spreading Dynamics in Quasiperiodic Non-Hermitian Systems}

\author{Ze-Yu Xing\textsuperscript{1, 2}}
\affiliation{\textsuperscript{1}Beijing National Laboratory for Condensed Matter Physics, Institute of Physics, Chinese Academy of Sciences, Beijing 100190, China}
\affiliation{\textsuperscript{2}School of Physical Sciences, University of Chinese Academy of Sciences, Beijing 100049, China}
\author{Shu Chen\textsuperscript{1, 2}}
\affiliation{\textsuperscript{1}Beijing National Laboratory for Condensed Matter Physics, Institute of Physics, Chinese Academy of Sciences, Beijing 100190, China}
\affiliation{\textsuperscript{2}School of Physical Sciences, University of Chinese Academy of Sciences, Beijing 100049, China}
\author{Haiping Hu\textsuperscript{1, 2}}\email{hhu@iphy.ac.cn}
\affiliation{\textsuperscript{1}Beijing National Laboratory for Condensed Matter Physics, Institute of Physics, Chinese Academy of Sciences, Beijing 100190, China}
\affiliation{\textsuperscript{2}School of Physical Sciences, University of Chinese Academy of Sciences, Beijing 100049, China}
%\date{\today}
\begin{abstract}
Non-Hermitian systems exhibit a distinctive type of wave propagation, due to the intricate interplay of non-Hermiticity and disorder. Here, we investigate the spreading dynamics in the archetypal non-Hermitian Aubry-André model with quasiperiodic disorder. We uncover counter-intuitive transport behaviors: subdiffusion with a spreading exponent $\delta=1/3$ in the localized regime and diffusion with $\delta=1/2$ in the delocalized regime, in stark contrast to their Hermitian counterparts (halted vs. ballistic). We then establish a unified framework from random-variable perspective to determine the universal scaling relations in both regimes for generic disordered non-Hermitian systems. An efficient method is presented to extract the spreading exponents from Lyapunov exponents. The observed subdiffusive or diffusive transport in our model stems from Van Hove singularities at the tail of imaginary density of states, as corroborated by Lyapunov-exponent analysis.
\end{abstract}
\maketitle
Anderson localization is a fundamental wave phenomenon originating from interference effect in disordered media \cite{AL1, AL2}. Typically, in Anderson localized systems, eigenstates are exponentially confined in space, resulting in a complete halt of wave propagation or particle transport, i.e., dynamical localization. Recently, non-Hermiticity has emerged as a pivotal ingredient \cite{coll1, coll4, coll6, colladd3, nhreview, nhreview2} in various platforms, including photonic, acoustic, cold atomic, and dissipative quantum systems. Non-Hermiticity can be introduced, e.g., via complex on-site potentials, reflecting energy or particle exchange with the environment. The interplay between disorder and non-Hermiticity gives rise to a wealth of intriguing phenomena like spectral localization \cite{nhat1, nhat2, nhat3, nhat4}, non-Hermitian mobility edge \cite{me1, me2, me3, me4, me5, me6, me7,me8}, topological Anderson insulator \cite{nhtai1, nhtai2, nhtai3, nhtai4, nhtai5}, and scale-free or tailored localization \cite{sfl1, sfl2, sfl3, sfl4,sfl5,stl}.

\begin{figure}[!t]
\centering
\includegraphics[width=3.33 in]{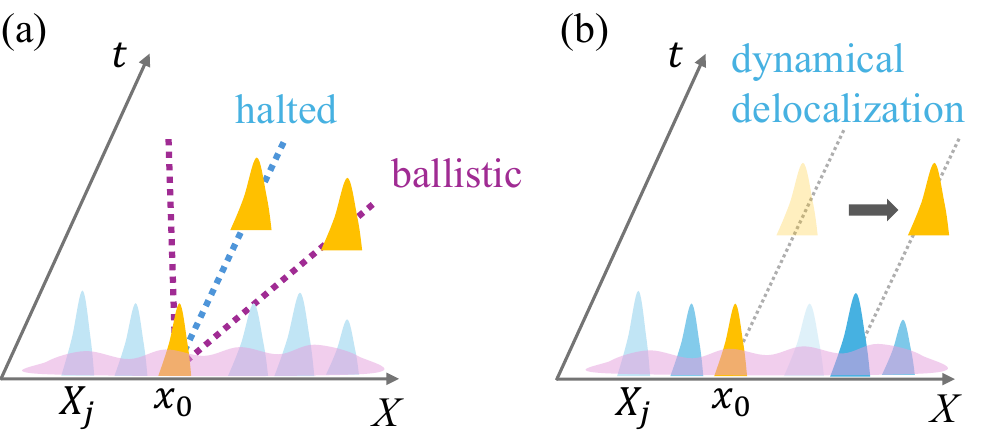}
\caption{Comparison of wave spreading in (a) Hermitian and (b) non-Hermitian lattices. Eigenstates are extensive (purple) in the delocalized regime and exponentially confined (blue) in the localized regime. In the Hermitian case, an initial wave packet (orange) spreads ballistically (dashed purple line) or remains localized near its starting point (dashed blue line) for the two regimes, respectively. In the non-Hermitian case, eigenstates with larger imaginary parts of energies (darker shades) dominate the evolution. The initial wave packet may ``jump” to other sites, leading to dynamical delocalization even in the localized regime.}\label{fig1}
\end{figure}

A scrutinization of Anderson localization in non-Hermitian systems led to a striking finding: in this setting, Anderson localization and dynamical localization may no longer align \cite{photonic_exp}. Non-Hermiticity introduces complex eigenenergies with imaginary components that modulate spatial distributions, causing wave profiles to ``jump”, as sketched in Fig. \ref{fig1}. In steady evolution, eigenstates with larger imaginary energy component dominate, enabling dynamical delocalization even in systems with fully localized eigenstates. The non-Hermiticity-induced jumpy dynamics has recently been observed in photonic lattices with engineered dissipation \cite{photonic_exp}. Such dynamical delocalization marks a key difference between unitary and nonunitary evolution, indicating that both the localization properties of eigenstates and spectral features shape the system’s dynamics. Yet, prior works has mainly focused on the localized regime \cite{photonic_exp,wz_dynamics} and the spreading exponents are determined on a case-by-case basis. To date, a unified framework governing the spreading dynamics in both regimes remains elusive. Key questions include: in the delocalized regime, does non-Hermiticity alter the ballistic transport typical of Hermitian systems? How is the wave spreading universally linked to spectral properties, and how to effectively extract the spreading exponents in a systematic way?

In this paper, we study the spreading dynamics in the non-Hermitian Aubry-André model. A key advantage of this model is that it hosts both delocalized and localized phases, with the Anderson transition analytically ascertained by Avila’s global theory \cite{avila}. We show that transport is diffusive with spreading exponent $\delta = 1/2$ in the delocalized regime and subdiffusive with $\delta = 1/3$ in the localized regime. We then formulate the jumpy dynamics of wave spreading and derive universal scaling relations for both regimes. The spreading exponent is shown to be linked to the imaginary density of states (iDOS) at the band tail and can be accurately extracted from Lyapunov exponents (LEs) in the complex plane. For our model, we attribute both the diffusive (in the delocalized regime) and subdiffusive transport (in the localized regime) to Van Hove singularities at the band tail. Our framework is applicable to generic disordered non-Hermitian systems, whether the disorder is correlated or uncorrelated, diagonal or off-diagonal, and Hermitian or non-Hermitian.

{\color{blue}\textit{Anderson transition.---}}We consider the following quasiperiodic non-Hermitian model, with Hamiltonian:
\begin{eqnarray}\label{model1}
H=\sum_j t(c_j^{\dag}c_{j+1}+c_{j+1}^{\dag}c_j)+V\cos(2\pi\alpha j+\phi)c_j^{\dag}c_j.
\end{eqnarray}
Here, the hopping strength $t=1$ is set as the energy unit, $\alpha=\frac{\sqrt{5}-1}{2}$ is the golden ratio and $\phi$ is a tunable phase. The onsite potential takes complex values, $V=|V|e^{i\phi_V}$. The case of $\phi_V=0,\pi$ reduces to the renowned Aubry-Andr\'{e} model \cite{AAmodel}, which undergoes Anderson transition from delocalized phase for weak disorder ($|V|<2t$) to localized phase for strong disorder ($|V|>2t$). This transition can be inferred from duality property in the Hermitian case. Unfortunately, when $V$ is complex, the self-duality is lost. The model (\ref{model1}) does not exhibit the non-Hermitian skin effect \cite{nhse1,nhse2} and features no spectral loops in the energy spectra \cite{nhse3,nhse4,nhse5,nhse6,nhse7,nhse8}. Unless otherwise noted, we take open boundary condition. The model satisfies $H\rightarrow H^*$ if $\phi_V\rightarrow -\phi_V$ and $H\rightarrow -H$ if $\phi_V\rightarrow\phi_V+\pi$, along with $c_j\rightarrow (-1)^{j}c_j$. Therefore, we restrict $\phi_V\in[0,\pi/2]$ and the spectra for other values of $\phi_V$ can be inferred from these transformations.

To pinpoint the transition point for the general case, we consider an eigenstate $|\psi\rangle=(\psi_1,\psi_2,\cdots)$ with energy $E$. The eigenvalue equation $t\psi_{j+1}+t\psi_{j-1}=[E-V\cos(2\pi\alpha j+\phi)]\psi_j$ can be recast into a transfer matrix form $(\psi_{j+1},\psi_j)=T_j(\psi_j,\psi_{j-1})$, with
\begin{eqnarray}
T_j(E)=\left(\begin{array}{cc}
 \frac{E-V\cos(2\pi\alpha j+\phi)}{t} & -1 \\
1 & 0\\
\end{array}\right).
\end{eqnarray}
The LE is defined as
\begin{eqnarray}\label{le}
\gamma(E)=\lim_{L\rightarrow\infty}\frac{1}{L}\ln \left\lVert\prod_{j=1}^L T_j\right\rVert,
\end{eqnarray}
with $||.||$ the matrix norm. By Avila's global theory \cite{avila,SM}, when $E$ is located inside the spectra, the LE is
\begin{eqnarray}\label{leavila}
\gamma(E)=\max\left\{\ln \left\lvert\frac{V}{2t}\right\rvert,0\right\}.
\end{eqnarray}
A positive value of $\gamma(E)$ signifies a localized eigenstate. The Anderson transition occurs when $\gamma(E)=0$, i.e., $|V|=2|t|$. Thus, the LE is irrelevant to the choice of $E$ within the spectra or the phases $\phi$, $\phi_V$. It means that all eigenstates undergo the transition simultaneously, with no mobility edge in this model.
\begin{figure}[!t]
\centering
\includegraphics[width=3.33 in]{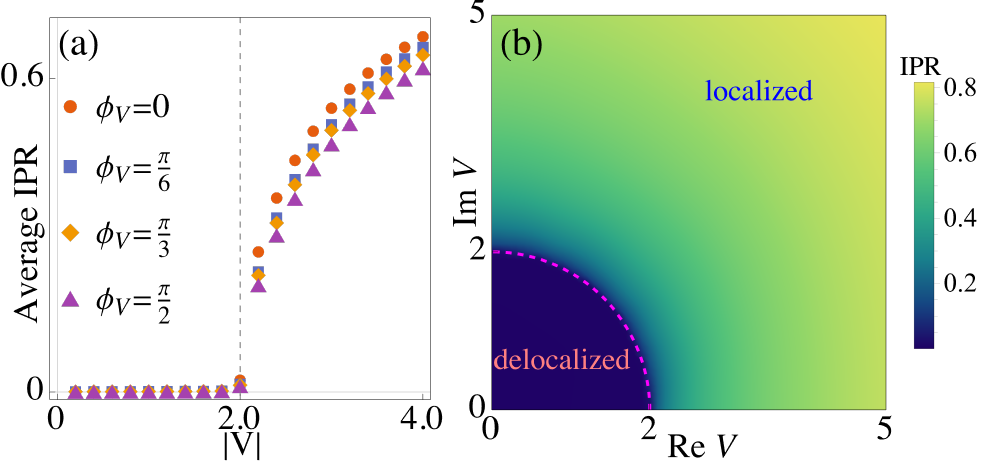}
\caption{Anderson localization transition in the quasiperiodic non-Hermitian model (\ref{model1}). (a) Average inverse partition ratio (IPR) versus disorder strength $|V|$ for four different phases $\phi_{V}=0,\pi/6,\pi/3$ and $\pi/2$. The vertical dashed line marks the transition point $|V|=2$. (b) Contour plot of the average IPR in the $(\textrm{Re}V, \textrm{Im}V)$ plane. The phase boundary $(\textrm{Re} V)^2+(\textrm{Im} V)^2=4$ (dotted line) separates the delocalized regime ($|V|<2$) from the localized regime ($|V|<2$). The system length is set to $L=2584$, $\phi=0$.}\label{fig2}
\end{figure}

The localization of eigenstates can be characterized by the inverse participation ratio (IPR). For an eigenstate $|\psi\rangle=(\psi_1,\psi_2,\cdots,\psi_L)$, with $L$ the system length, the IPR is defined as
\begin{eqnarray}
\textrm{IPR}=\sum_{n=1}^{L} |\psi_n|^4.
\end{eqnarray}
For extended states, the IPR approaches zero, whereas for localized states, it takes a finite value. In Fig. \ref{fig2}(a), we plot the average IPR over all eigenstates as a function of the disorder strength $|V|$ for four representative phase factors in the quasiperiodic potential, $\phi_V=0, \pi/6, \pi/3, \pi/2$. The IPR remains zero in the extended phase ($|V|<2$) and rises to a finite value in the localized phase ($|V|>2$). A contour plot of the average IPR in the $(\textrm{Re}V,\textrm{Im}V)$ space is shown in Fig. \ref{fig2}(b), with a clear phase boundary at $(\textrm{Re} V)^2+(\textrm{Im} V)^2=4$ separating the two regimes. These numerical results agree with the transition point predicted by Avila's global theory. It is noteworthy that the conclusion also holds for left eigenstates and all $\phi_V\in[0,2\pi]$ in our model.

{\color{blue}\textit{Dynamical spreading.---}}We now examine the spreading dynamics of a wave packet on this quasiperiodic lattice. In non-Hermitian systems, the time evolution is non-unitary, and the total probability is generally not conserved. For an initial excitation $|\Psi_0\rangle$ (e.g., placed at site $x_0$ far from the boundary), the time-evolved state is given by $|\Psi(t)\rangle = e^{-iHt}|\Psi_0\rangle$. In the site basis, $|\Psi(t)\rangle = \sum_j \psi_j(t) |j\rangle \langle j|$. A normalization procedure is imposed such that $\sum_j |\psi_j(t)|^2 = 1$ holds at all times. Physically, this corresponds to disregarding non-detection events in the dynamics and is relevant in, e.g., light propagation in photonic lattices \cite{photonic_exp} or discrete-time quantum walks \cite{xuepeng1,xuepeng2,xuepeng3}. The wave spreading is quantified by the second moment of wave-packet distribution:
\begin{eqnarray}
X^2(t)=\sum_j j^2|\psi_j(t)|^2. 
\end{eqnarray}
We consider an ensemble of time-evolved state $\{|\Psi(t)\rangle\}$, with varying the tuning phase $\phi$ and initial locations $x_0$ (provided the far-from-boundary condition is met). The ensemble average of the deviation from initial positions gives the spreading exponent $\delta$, i.e.,
\begin{eqnarray}
X(t)\sim t^{\delta},
\end{eqnarray}
where $X(t) = \langle\sqrt{X^2(t)}\rangle_{ave}$. $\delta$ is the slope of $\frac{d \log X(t)}{d \log t}$ during steady evolution, where $\delta > 1/2$, $\delta = 1/2$, and $\delta < 1/2$ corresponds to superdiffusive, diffusive, and subdiffusive transport, respectively.
\begin{figure}[!t]
\centering
\includegraphics[width=3.33 in]{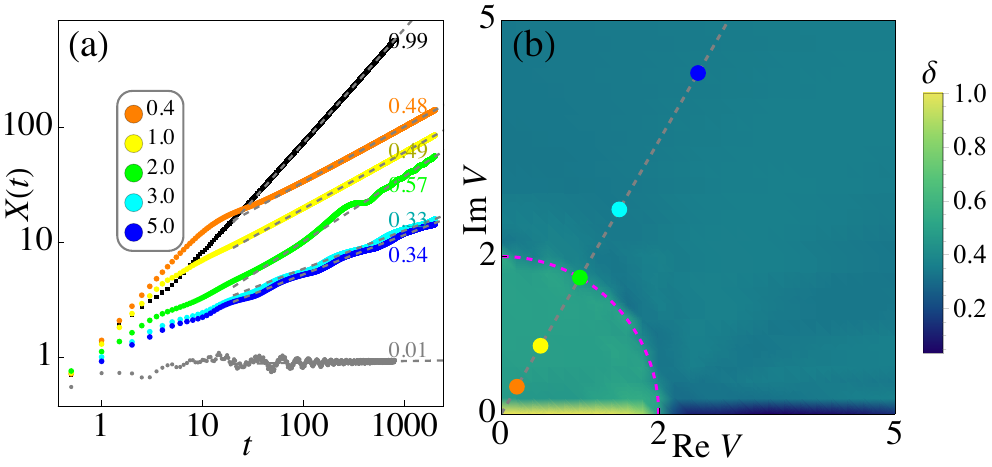}
\caption{Spreading dynamics of model (\ref{model1}). (a) Average deviation $X(t)$ from the initial position versus time $t$ for various disorder strengths $|V|$. $\phi_V = \pi/3$. The spreading exponent $\delta$ is extracted by linear fitting the slope during steady evolution (dashed lines). The ensemble average is over dozens of phase samples of $\phi$ and initial positions $x_0$. For comparison, the Hermitian case ($\phi_V=0$) is shown for $V=1$ (black dots) and $V=4$ (gray dots). (b) Contour plot of $\delta$ in the $(\textrm{Re}V, \textrm{Im}V)$ plane. The colored dots correspond to the parameter selections in (a). The dashed line marks the phase boundary. On the real axis (the Hermitian case), the transport is ballistic or impeded. The system length is set to $L=2584$.}\label{fig3}
\end{figure}

In Fig. \ref{fig3}(a), we show $X(t)$ versus $t$ for model (\ref{model1}) across different parameter regimes. After initial transient dynamics, the wave spreading stabilizes, and a numerical fitting of the slope is performed during steady evolution. For the Hermitian case ($\phi_V = 0$), which was extensively studied in the literature, the numerics confirm ballistic transport ($\delta \approx 1$) in the delocalized phase and impeded transport ($\delta \approx 0$) in the localized phase. In sharp contrast, the non-Hermitian case ($\phi_V = \pi/3$) exhibits markedly different behavior. In the delocalized regime, the numerical results yield, e.g., $\delta = 0.48$ for $|V| = 0.4$ and $\delta = 0.49$ for $|V| = 1$. Whereas in the localized regime, $\delta = 0.33$ for $|V| = 3$ and $\delta = 0.34$ for $|V| = 5$. At the transition point $|V| = 2$, numerical fitting gives $\delta = 0.57$. In Fig. \ref{fig3}(b), we present the dynamical phase diagram in the $(\textrm{Re}V,\textrm{Im}V)$ plane. It is clear that the spreading appears universal in each regime: for the non-Hermitian case, $\delta \approx 1/2$ in the entire delocalized phase and $\delta \approx 1/3$ in the entire localized phase. 

{\color{blue}\textit{Universal scaling relations.---}}To understand the distinct spreading behaviors in different regimes, we analyze the propagator on a generic $d$ dimensional non-Hermitian lattice. Consider an initial wave packet placed at the central site $\bold{x}_0=\bold{0}$. In the fully localized regime, with a typical localization length $\xi$, an eigenstate of energy $E$ is associated with its localization center $\bold{X}$, denoted as $E(\bold{X})=\epsilon(\bold{X})+i\lambda(\bold{X})$. In the thermodynamic limit and under ensemble average, the eigenspectra follow certain distributions in the complex-energy plane. The specific forms of $\epsilon(\bold{X})$ and $\lambda(\bold{X})$ depend on the model and type of disorder. The propagator is
\begin{eqnarray}
\langle \bold{X}|e^{-iHt}|\bold{x}_0\rangle \sim e^{-i\epsilon(\bold{X}) t}e^{\lambda(\bold{X}) t - \frac{X}{\xi}},
\end{eqnarray}
with $X=|\bold{X}-\bm{x}_0|$. The real part $\epsilon(\bold{X})$ of the eigenenergy contributes purely a dynamical factor. In the second exponential, there are two competing terms: $e^{-X/\xi}$ accounts for the overlap between $|\bold{X}\rangle$ and the initial wave packet; while the $e^{\lambda(\bold{X})t}$ term modulates the wave packet's distribution in the eigenbasis. Over time, the dominant eigenstate is dictated by the maximization condition \cite{wz_dynamics}:
\begin{eqnarray}\label{rel1}
\frac{\partial\lambda}{\partial X} \sim t^{-1}.
\end{eqnarray}

In the delocalized regime, a Fourier transformation is applied to convert extended states in real space into localized states, with a typical localization length $\tilde{\xi}$ in the reciprocal space. An eigenstate with energy $E(\bold{K})=\epsilon(\bold{K})+i\lambda(\bold{K})$ is characterized by its localization center $\bold{K}$. The propagator is
\begin{eqnarray}
\langle \bold{X}|e^{-iHt}|\bold{x}_0\rangle \sim \int d\bold{K} e^{-i\epsilon(\bold{K}) t} e^{\lambda(\bold{K}) t} \langle \bold{X}|\bold{K}\rangle \langle \bold{K}|\bold{x}_0\rangle.
\end{eqnarray}
As $\langle \bold{X}|\bold{K}\rangle = \int d \bold{k} e^{i\bold{kX} - \frac{|\bold{k-K}|}{\tilde{\xi}}} \sim (X^2+\tilde{\xi}^{-2})^{-\frac{d+1}{2}}\sim X^{-d-1}$ for large $X$, the maximization condition yields
\begin{eqnarray}\label{rel2}
X\frac{\partial \lambda}{\partial X} \sim t^{-1}.
\end{eqnarray}

From above, it is evident that the imaginary parts of the eigenspectra play a crucial role. Let $\rho_I(s)$ denote the normalized distribution (i.e., $\int_{-\infty}^{\infty} d s \rho_I(s) = 1$) of the imaginary parts of the spectra in the thermodynamic limit, i.e., the iDOS, averaged over the ensemble. The evolution from $\bold{x}_0$ to $\bold{X}(t)$ is governed by a local Hamiltonian of size $X^d$. Note that the wave packet reaches the ``boundary" at time $t$, $\lambda$ in Eqs. (\ref{rel1})(\ref{rel2}) thus corresponds to the largest imaginary part among all $O(X^d)$ eigenvalues of this local Hamiltonian. A simple calculation of the cumulative distribution immediately gives \cite{SM}:
\begin{eqnarray}\label{rel3}
\int_{\lambda}^{\infty}d s\rho_I(s)\sim X^{-d},
\end{eqnarray}
under ensemble average. By combining Eqs. (\ref{rel1})(\ref{rel3}) (or Eqs. (\ref{rel2})(\ref{rel3})), the scaling relation for steady evolution is determined in the localized (or delocalized) regime. Notably, the spreading exponent $\delta$ depends solely on the behavior of iDOS near the band tail. For instance, when the iDOS scales algebraically at the tail, $\rho_I(s) \sim (s_0 - s)^{\beta} \Theta(s_0 - s)$, the spreading exponents are given by
\begin{eqnarray}\label{scalingrelation}
\delta = \frac{\beta + 1}{d+\beta + 1}; ~~\delta=\frac{\beta+1}{d},
\end{eqnarray}
for the localized and delocalized regimes, respectively.

\begin{figure}[!t]
\centering
\includegraphics[width=3.33 in]{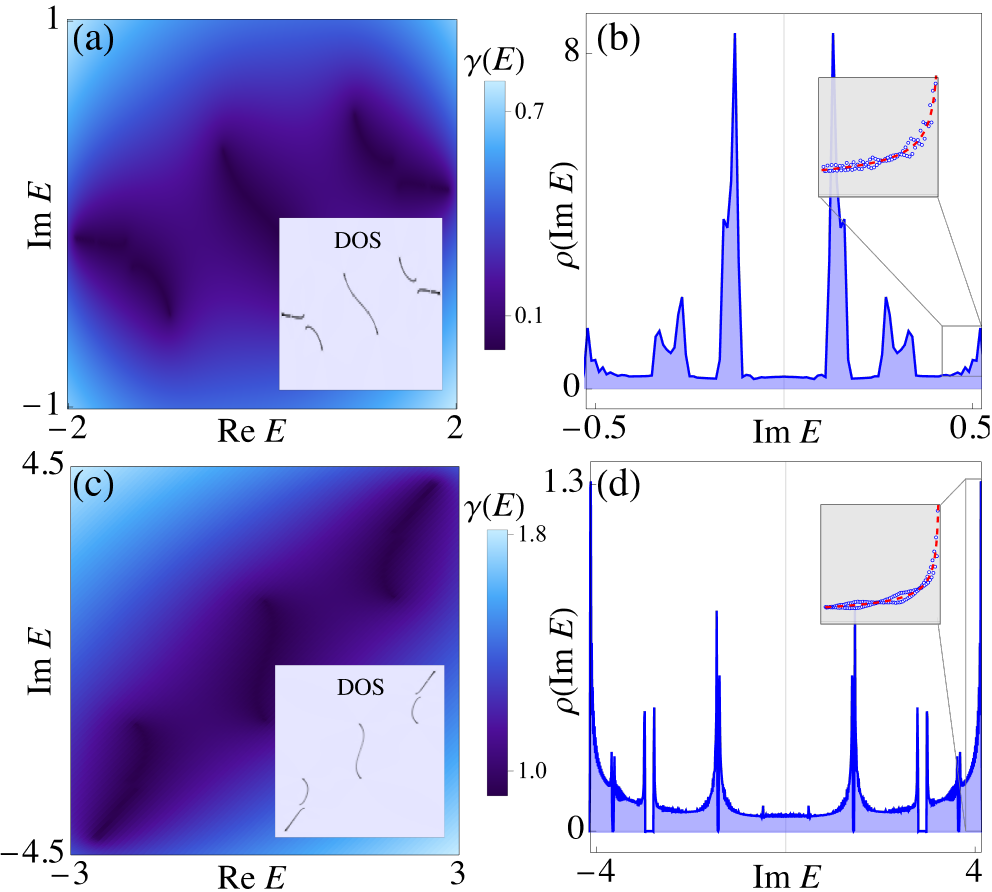}
\caption{Lyapunov exponent (LE) analysis of the spreading exponent. (a)(c) Contour plot of the LE $\gamma(E)$ in the complex plane for $|V|=1$ (delocalized regime) and $|V|=5$ (localized regime), respectively. (Inset) The Laplacian of $\gamma(E)$ yields the density of states. (b)(d) The iDOS extracted from the LE [See Eqs. (\ref{idos1})(\ref{idos2}))]. The divergence at the band tail is fitted (inset) as $\rho_I(\lambda)=0.023+\frac{0.088}{\sqrt{0.525-\lambda}}$ for (b) and $\rho_I(\lambda)=0.022+\frac{0.099}{\sqrt{4.13-\lambda}}$ for (d). For all panels, $\phi_V=\pi/3$.}\label{fig4}
\end{figure}
{\color{blue}\textit{Analysis of LE.---}}In generic disordered non-Hermitian systems, determining the exponent $\delta$ boils down to evaluating the iDOS at the band tail. In many cases, this can be handled by fitting the numerical energy spectra, but it is not always an easy task. In fact, achieving the thermodynamic limit is challenging, and exact diagonalization of large non-Hermitian Hamiltonians can be prone to severe numerical errors \cite{error1,nhse8}. Even worse, precision cannot be guaranteed when divergences exist in the iDOS. Here, we present a powerful approach to extract the exponent $\delta$. It begins by analytically continuing of the LE [See Eq. (\ref{le})] into the whole complex plane in the transfer matrix formalism. The density of states (DOS) relates to the LE via the generalized Thouless relation \cite{thouless1,gtr}:
\begin{eqnarray}\label{idos1}
\rho(E)=\frac{1}{2\pi}\nabla^2\gamma(E).
\end{eqnarray}
As key advantages, this method avoids diagonalizing large non-Hermitian matrices or performing numerous ensemble average. (The transfer-matrix algorithm only involves multiplying small matrices.) In certain cases, it even allows for analytical treatments. By integrating out the real part of the eigenenergies, we obtain the iDOS:
\begin{eqnarray}\label{idos2}
\rho_I(\textrm{Im} E)=\int \rho(E)~d\textrm{Re} E.
\end{eqnarray}

For our model (\ref{model1}), the profiles of the LE in the complex plane for $|V|=1$ (delocalized) and $|V|=5$ (localized) are shown in Figs. \ref{fig4}(a) and (c), respectively. The Laplacian of the LE yields the DOS [See inset]. The corresponding iDOS, obtained by integrating out the real components, is shown in Figs. \ref{fig4}(b) and (d). Several spectral peaks are visible, with a divergence at the band tail dictating the spreading dynamics. Numerical fitting near the tail reveals that, in both regimes, the peak corresponds to a Van Hove singularity with $\beta = -1/2$. According to Eq. (\ref{scalingrelation}), the spreading exponents are $\delta = 1/3$ in the localized phase and $\delta = 1/2$ in the delocalized phase, consistent with the scaling relations in Fig. \ref{fig3}. We note that the singularity in the iDOS stems from the quasiperiodic potential in model (\ref{model1}). A perturbative analysis in the deep localized regime \cite{SM} suggests the stability of $\beta = -1/2$ divergence. 

{\color{blue}\textit{Conclusion and discussions.---}}To conclude, we investigated the non-Hermitian Aubry-André model with quasiperiodic disorder and identified counter-intuitive subdiffusive ($\delta = 1/3$) and diffusive ($\delta = 1/2$) wave spreading in the localized and delocalized regimes, respectively. By linking the anomalous dynamical delocalization to nonunitary jumps of wave profiles, the universal scaling relations were obtained to determine the spreading exponent $\delta$ for both regimes. We also introduced an efficient method to extract the exponent $\delta$ from the analysis of LEs. Our findings reveal the sharp distinction between Hermitian and non-Hermitian dynamics and pave the way for studying wave spreading in generic disordered non-Hermitian systems.

Our results highlight the intriguing interplay between non-Hermiticity and disorder. For other models or disorder types, the spreading exponent can be determined in a streamlined manner via Eqs. (\ref{idos1})(\ref{idos2}) and the scaling relations Eqs. (\ref{rel1})(\ref{rel2})(\ref{rel3}). In non-Hermitian systems, disorder may lead to the emergence of mobility edges \cite{me1, me2, me3, me4, me5, me6, me7}. Since jumpy dynamics selectively favor certain eigenmodes, the wave spreading should remain to be dominated by eigenstates near the band tail \cite{SM}. A critical scenario arises when the mobility edge coincides with the band tail or when the system is at the transition point, such as $|V| = 2t$ in our model, where all eigenstates exhibit multifractality. Determining the spreading exponent in such critical cases remains an open question. Additionally, the wave spreading should be irrelevant to boundary conditions, as long as steady-state evolution is considered and the boundary is not reached. With advances in engineering various types of non-Hermiticity and disorder in platforms like photonic, acoustic, cold atomic, and dissipative quantum systems, the universal spreading dynamics should be readily observed in these highly tunable systems.

\begin{acknowledgments}
This work is supported by the National Key Research and Development Program of China (Grants No. 2023YFA1406704 and No. 2022YFA1405800) and National Natural Science Foundation of China (Grant No. 12474496). S. C. is also supported by the NSFC (Grants No. 12174436, No. T2121001, and No.12474287) and the Strategic Priority Research Program of Chinese Academy of Sciences under Grant No. XDB33000000.
\end{acknowledgments}

% Define new numbering format for supplementary equations
\renewcommand{\theequation}{S\arabic{equation}}
\setcounter{equation}{0}  % Reset the equation counter

\clearpage
\appendix
\renewcommand{\thefigure}{S\arabic{figure}}
\setcounter{figure}{0}
\pagebreak
\widetext
\begin{center}
\textbf{\large  Supplemental Material}
\end{center}

This supplemental material provides additional details on
	
(I) Lyapunov exponent inside the energy spectra;

(II) Derivation of Eq. (\ref{rel3}) in the main text;
	
(III) Perturbative analysis of the spectral structure;

(IV) Dynamical spreading in the presence of mobility edge.

\section{(I)~Lyapunov exponent inside the energy spectra}\label{appendix1}
In this appendix, we calculate the Lyapunov exponent (LE) inside the energy spectra [See Eq. (\ref{leavila}) in the main text]. We start with the eigenvalue equation
\begin{align}
t (\varphi_{j-1} + \varphi_{j+1} ) + V_{j} \varphi_{j} = E \varphi_{j},
\end{align}
where $V_{j} = \left| V \right|e^{ i \phi_{V} }\cos(2\pi \alpha j + \phi)$. The transfer matrix is written as
\begin{align}
T_{j}(E) = \begin{pmatrix}
\frac{E - V_{j}}{t} & -1 \\
1 & 0
\end{pmatrix}.
\end{align}
The LE is defined as 
\begin{align}
\gamma(E) = \lim_{ L \to \infty } \frac{1}{L}\ln \left\lVert  \prod_{j=1}^{L} T_{j}(E) \right\rVert.
\end{align}
To calculate $\gamma(E)$, we perform an analytical continuation of the phase $\phi \to \phi + i \epsilon$ in $T_{j}$, i.e.,
\begin{align}
T_{j}(\epsilon; E) = \begin{pmatrix}
\frac{E}{t} - \frac{\left| V \right| e^{ i \phi_{V} }}{2t}[e^{-\epsilon + i(2\pi \alpha j + \phi)} + e^{ \epsilon -i (2\pi \alpha j + \phi) }] & -1  \\
1 & 0
\end{pmatrix}.
\end{align}
In the large-$\epsilon$ limit, only the term with $e^{ \epsilon }$ survives:
\begin{align}
T_{j}(\epsilon; E) = \begin{pmatrix}
-\frac{\left| V \right| e^{ i \phi_{V} }}{2t}e^{ \epsilon - i (2\pi \alpha j + \phi ) } & 0 \\
 0 & 0
\end{pmatrix} + \mathcal{O}(1).
\end{align}
Accordingly,
\begin{align}
\gamma_{\epsilon}(E) = \lim_{ L \to \infty } \frac{1}{L} \ln \left\lVert  \prod_{j=1}^{N} T_{j}(\epsilon; E) \right\rVert = \epsilon + \ln \left\lvert  \frac{V}{2t}  \right\rvert .
\end{align}
Avila's global theory \cite{avila} tells that, as a function of $\epsilon$, $\gamma_{\epsilon}(E)$ is a convex, piecewise linear function, whose slopes are integers. It follows that $\gamma_{\epsilon}(E) = \max \left\{ \epsilon + \ln \left\lvert  \frac{V}{2t}  \right\rvert, \gamma_{0}(E) \right\}$. $\gamma_{\epsilon}(E)$ is an affine function in a neighborhood of $\epsilon=0$. Moreover, an energy does not belong to the spectra, if and only if $\gamma_{0}(E) > 0$. As a result, we have
\begin{align}
\gamma(E) = \max \left\{ \ln \left\lvert  \frac{V}{2t}  \right\rvert , 0 \right\}
\end{align}
for any energy $E$ lies inside the spectra. 

\section{(II)~Derivation of Eq. (\ref{rel3}) in the main text}
Let us revisit the basic setup. We consider a non-Hermitian Hamiltonian of size $N=X^d$ ($d$ is the spatial dimension), whose imaginary parts of eigenvalues are labeled as $\lambda_1,\lambda_2,\cdots,\lambda_N$. In the thermodynamic limit and under ensemble averaging, the imaginary part obeys a normalized distribution $\rho_I(s)$, $\int ds \rho_I(s)=1$. Let us denote $\lambda_{max}$ as the largest among all $\lambda_j$'s. Note that $\lambda_{max}$ by itself is a random variable. By treating $\lambda_j~(j=1,2,\cdots,N)$ as random variables drawn from the distribution $\rho_I(s)$, we need to show that ($\mathbb{E}[.]$ means ensemble average)
\begin{align}
\mathbb{E}[\int_{\lambda_{\max}}^{\infty} d s \rho_{I}(s)] \sim \frac{1}{N}.
\end{align}

To this end, we consider the cumulative distribution function:
\begin{align}
F(\lambda) = \int_{-\infty}^{\lambda} d s \, \rho_{I}(s),
\end{align}
which gives the probability that in a single realization, the imaginary component is less than $\lambda$. The probability with $\lambda_{max}\leq \lambda$ is
\begin{align}
P(\lambda_{\max}\leq \lambda) = \left[ F(\lambda) \right] ^{N}.
\end{align}
The probability distribution function (PDF) of $\lambda_{\max}$ (denoted as $p_{max}$) is the derivative of the above cumulative distribution function:
\begin{align}
p_{\max}(\lambda) = \frac{d}{d \lambda} P(\lambda_{\max} \leq \lambda) = N \left[ F(\lambda) \right] ^{N-1} \rho_{I} (\lambda).
\end{align}
We thus have the expectation value (under ensemble average)
\begin{eqnarray}
\mathbb{E} \left[\int_{\lambda_{\max}}^\infty d \lambda \, \rho_{I}(\lambda) \right]= \mathbb{E} [1 - F(\lambda_{\max})]= \int_{-\infty}^{\infty} d \lambda \, p_{\max}(\lambda)\left[ 1 - F(\lambda) \right]= \int_{-\infty}^{\infty} d \lambda  \, N\left[ F(\lambda)\right]^{N-1} \rho_{I}(\lambda) \left[ 1 - F(\lambda) \right].
\end{eqnarray}
Note that $\rho_{I}(\lambda)d \lambda = dF(\lambda)$ and replace $F({\lambda})$ by $u$ with $u \in [0,1]$, we have
\begin{align}
\mathbb{E} \left[\int_{\lambda_{\max}}^\infty d \lambda \, \rho_{I}(\lambda) \right]   = N \int_{0}^{1} du \, u^{N-1}(1-u)  = 1 - \frac{N}{N+1}  = \frac{1}{N+1}
\end{align}
Consequently, in the large $N$ limit, the expectation value is $1/N$ and Eq. (\ref{rel3}) is proved.

\section{(III)~Perturbative analysis of the spectral structure}
Our Hamiltonian contains two pars: 
\begin{align}
H = H_{V} + H_{t} = \begin{pmatrix}
V \cos (2\pi \alpha + \phi) & t   \\
t & \ddots  & \ddots  &  \\
 & \ddots  & V \cos(2\pi \alpha j + \phi ) & t \\
 &  & t  &  \ddots  & \ddots  &  \\
 &  &  & \ddots  &  &  t\\
 &  &  &  & t &  V \cos (2\pi \alpha L + \phi)
\end{pmatrix},
\end{align}
where $H_V$ is the (diagonal) onsite potential with $V = \left| V \right|e^{ i \phi_{V} }$ and $H_t$ is the hopping term. For convenience, the length of the lattice is set as a Fibonacci number $L = F_{N}$. When $t=0$, $H = H_{V}$ is already diagonalized. We first show that in the thermodynamic limit, the spectra form a line parametrized by a real wave number $k$ \cite{yangchao} uniformly drawn from $-\pi$ to $\pi$, i.e., $E(k) = V \cos k$ with $k \in [-\pi, \pi)$. The eigenenergies are $E_{j}^{(0)} = V \cos (2\pi \alpha j + \phi)$. We rewrite $\alpha$ as
\begin{align}
\alpha = \frac{F_{N-1}}{F_{N}} + \beta_{N} \equiv \alpha_{N} + \beta_{N}.
\end{align}
We have $E_{j}^{(0)} = V \cos \left( \frac{2\pi}{F_{N}} F_{N-1}j +  2\pi \beta_{N}j + \phi \right)$. Using $F_{N} = \frac{1}{\sqrt{ 5 }}[\alpha^{-N} + (-\alpha)^{N}]$, we have $\beta_{N} \sim \frac{1+\alpha^{2}}{\alpha}(-1)^{N+1}\alpha^{2N}$ as $N\to \infty$. Then $\left| \beta_{N}j \right| \leq \left| \beta_{N}F_{N} \right| \sim \alpha^{N} \to 0$ as $ N\to \infty$. Thus the $2\pi \beta_{N}j$ term can be omit in the thermodynamic limit. We introduce $n = jF_{N-1} \mod{F_{N}}$ for $j = 1, 2, \dots, F_{N}$. It is easy to verify that $n = 0, 1, \dots, F_{N} -1$. Let $k = \frac{2\pi}{F_{N}}n$, we can label the eigenenergy as $E^{(0)}(k) = V \cos(k + \phi)$ for $k \in [0, 2\pi)$. As the phase $\phi$ is irrelevant in the spectral structure, it can be omitted. The imaginary component of the spectra $\lambda= \mathrm{Im}E = \left| V \right| \sin \phi_{V} \cos k$ follows the distribution 
\begin{align}
\rho_I(\lambda) = \frac{1}{\left| d \lambda / dk \right| } = \frac{1}{\left| V \sin \phi_{V} \sin k\right| } = \frac{1}{\sqrt{ \left| V \right| ^{2} \sin ^{2}\phi_{V} - \lambda^{2} }}\sim (\lambda_{max}-\lambda)^{-1/2},
\end{align}
with $\lambda_{max}=|V|\sin^2\phi_V$. Thus, there is a Van Hove singularity at the band tail with $\beta=-1/2$. In the following, we show the stability of this singularity under perturbation. We consider $t \ll \lvert V \rvert$ and treat $H_{t}$ as perturbation. It is easy to check that the first-order perturbation vanishes. The second-order perturbation can be computed directly:
\begin{align}
E_{j}^{(2)} = - \sum_{l\neq j} \frac{\left|\bra{\psi_{l}^{(0)}}H_{t}\ket{\psi_{j}^{(0)}}\right|^{2}}{E_{l}^{(0)} - E_{j}^{(0)}} = \frac{t^{2}}{V} \frac{\cos(2\pi \alpha j + \phi)}{\cos ^{2}(2\pi \alpha j + \phi)- \cos ^{2}(\pi \alpha)}.
\end{align}
Similarly, the eigenenergy in the thermodynamic limit is labeled by $E(k)$, whose imaginary part is  
\begin{align}
\lambda = \mathrm{Im} E(k) = \left| V \right| \sin \phi_{V} \cos k \left( 1 - \frac{t^{2}}{\left| V \right|^{2}} \frac{1}{\cos ^{2}k-\cos ^{2}\pi \alpha} \right).
\end{align}
Note that only the spectral feature at the band tail is relevant for spreading dynamics. By replacing $\left| V \right|\sin \phi_{V} \cos k$ with $\lambda$, the distribution of the imaginary parts of eigenenergies is 
\begin{eqnarray}
\rho_I(\lambda) &&= \frac{1}{\left| d \lambda / dk \right| } = \frac{1}{\sqrt{ \left| V \right| ^{2} \sin ^{2} \phi_{V} - \lambda^{2}}}\left( 1 -  t^{2}\sin ^{2} \phi_{V} \frac{\lambda^{2} + \left| V \right| ^{2} \sin ^{2} \phi_{V} \cos ^{2}\pi \alpha}{(\lambda^{2} - \left| V \right| ^{2} \sin ^{2} \phi_{V} \cos ^{2}\pi \alpha)^2} \right)\notag\\
&&\sim (\lambda_{max}-\lambda)^{-1/2}[1+O(t^2)],
\end{eqnarray}
near the band tail. It is easy to see from the above analysis, the Van Hove singularity with $\beta=-1/2$ divergence at the band tail is stable against perturbation.

\section{(III)~Dynamical spreading in the presence of mobility edge}
\begin{figure*}[!t]
\centering
\includegraphics[width=6.5 in]{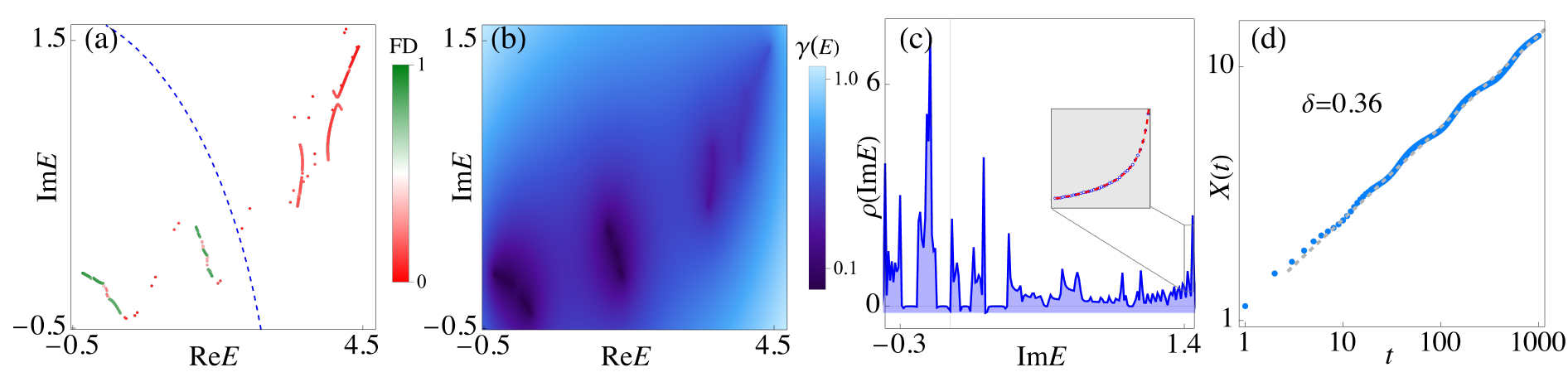}
\caption{Lyapunov exponent (LE) analysis of the spreading exponent for model (\ref{model2}). (a) Numerical energy spectra in the complex-energy plane, with colors denoting the fractal dimension (FD). The mobility edge (dotted blue line) is given by Eq. (\ref{meeq}). Ten evenly chosen $\phi$ inside $[0,2\pi]$ are included. The system length is set as $L = 987$. (b) Contour plot of the LE $\gamma(E)$ in the complex plane. (c) The iDOS extracted from the LE. The divergence at the band tail is fitted (inset) by Eq. (\ref{fitting2}). (d) The average deviation from initial position $\langle X(t)\rangle_{ave}$ versus time $t$. The slope in the steady evolution is $\delta=0.36$. For all panels, $t=1$, $b = 0.5$, $h = 1.5$, and  $V= e^{ i\pi/3 }$.}\label{figs}
\end{figure*}

In this section, we consider the following Hamiltonian \cite{me7}:
\begin{eqnarray}\label{model2}
H=\sum_j t(c_j^{\dag}c_{j+1}+c_{j+1}^{\dag}c_j)+V_j c_j^{\dag}c_j,
\end{eqnarray}
where 
\begin{align}
V_{j} = \frac{V \cos(2 \pi \alpha j + \phi) + h}{1-b \cos(2 \pi \alpha j + \phi)}.
\end{align}
Compared to model (\ref{model1}) in the main text, this model hosts mobility edge, which can be determined via Avila's global theory \cite{me7,avila}
\begin{align}\label{meeq}
\left( \mathrm{Re}E + \frac{\mathrm{Re}V}{b} \right)^{2} + \frac{\left( \mathrm{Im}E + \frac{\mathrm{Im}V}{b} \right)^{2}}{1-b^{2}} = \frac{4t^{2}}{b^{2}}.
\end{align}
In the following, we take parameter $t=1$, $b = 0.5$, $h = 1.5$, and  $V= e^{ i\pi/3 }$ as example. The localization property of an eigenstate can be characterized by its fractal dimension (FD), defined as:
\begin{eqnarray}
\textrm{FD}=-\frac{\ln\left( \sum_{x} \lvert \psi(x) \rvert^{4} \right)}{\ln L}.
\end{eqnarray}
In Fig. \ref{figs}(a), we show the numerical energy spectra (for 10 evenly chosen $\phi$ inside $[0,2\pi])$, with colors representing the FD. One can clearly observe two types of eigenstates separated by the mobility edge [Eq. (\ref{meeq})]. For extended or localized states, $\textrm{FD}\approx 1$ or $0$. The band tail of the imaginary parts corresponds to localized states. In Fig. \ref{figs}(c), we show the LE in the complex-energy plane calculated from transfer matrix, the Laplacian of which yields the spectral density in the thermodynamic limit. In Fig. \ref{figs}(b), we show the iDOS extracted from the LE. A numerical fitting of the singularity at the band tail yields $\beta=-0.434538$, with the fitted distribution:
\begin{align}\label{fitting2}
\rho_I(s)=-0.0980738 + \frac{0.163777}{(1.4574 - s)^{0.434538}}.
\end{align}
According to our scaling relation [See Eq. (\ref{scalingrelation}) in the main text], the spreading exponent is $\delta=\frac{\beta+1}{\beta+2}=0.3612$. This is confirmed in Fig. \ref{figs}(d), where we show the average deviation from initial position $\langle X(t)\rangle_{ave}$ versus time $t$. The average is performed over 11 initial positions and 200 different phases of $\phi$. The spreading exponent extracted from slope is $\delta=0.36$. 
\end{document}